\documentstyle{mn}
\def\bigstrut{\vrule width0pt height0.4truecm}
\font\japit = cmti10 at 10truept
\title[Measuring the cosmological constant with redshift surveys]
{\vglue-3.0truecm
\centerline{\japit Accepted for publication in Monthly Notices of the R.A.S.}
\vglue 2.5truecm\noindent
Measuring the cosmological constant with redshift surveys}
\author[W. E. Ballinger, J. A. Peacock and A. F. Heavens]
{W. E. Ballinger$^1$, J. A. Peacock$^2$ and A. F. Heavens$^1$\\
$^1$ \bigstrut Institute for Astronomy, University of Edinburgh,
Blackford Hill, Edinburgh EH9 3HJ, UK\\
$^2$ Royal Observatory,
Blackford Hill, Edinburgh EH9 3HJ, UK}

\newcommand{\longeq}[2]{
\begin{array}{l}
\displaystyle\!\!\!
#1 \phantom{\int} \\
\displaystyle\;\;\;\;
#2 \phantom{\int}
\end{array}
}

\newcommand{\longeqtwo}[2]{
\begin{array}{l}
\displaystyle\!\!\!
#1 \phantom{\int} \\
\displaystyle\!\!\!
#2 \phantom{\int}
\end{array}
}

\newcommand{\etal}{et al.}

\newcommand{\bd}{\begin{equation}}
\newcommand{\ed}{\end{equation}}
\newcommand{\be}{\begin{equation}}
\newcommand{\ee}{\end{equation}}
\newcommand{\bea}{\begin{eqnarray}}
\newcommand{\eea}{\end{eqnarray}}
\newcommand{\bean}{\begin{eqnarray}}
\newcommand{\eean}{\end{eqnarray}}

\newcommand{\pr}{^{\prime}}
\newcommand{\om}{\Omega_{m}}
\newcommand{\ov}{\Omega_{\Lambda}}
\newcommand{\lam}{\Lambda}
\newcommand{\smk}{{\mbox {\scriptsize\bf k}}}
\newcommand{\bk}{{\mbox {\bf k}}}
\newcommand{\smr}{{\mbox {\scriptsize\bf r}}}
\newcommand{\kpp}{k_{\perp}}
\newcommand{\kpl}{k_{\parallel}}
\newcommand{\fpp}{f_{\perp}}
\newcommand{\fpl}{f_{\parallel}}

\def\bib{\parskip=0pt\par\noindent\hangindent\parindent
    \parskip =2ex plus .5ex minus .1ex}

\newcommand{\ie}{i.e.}

\newcommand{\nn}{\nonumber \\}

\newcommand{\apj}{{ApJ}}

\newcommand{\mn}{{MNRAS}}
\newcommand{\na}{{Nature}}

\newcommand{\bfig}{\begin{figure}}
\newcommand{\efig}{\end{figure}}
\newtheorem{figudumx}{Figure}
\newenvironment{figdumy}{\begin{figudumx}\rm}{\end{figudumx}}
\newcommand{\bfg}{\begin{figdumy}}
\newcommand{\efg}{\end{figdumy}}

\newcommand{\gs}{\mathrel{\raise1.16pt\hbox{$>$}\kern-7.0pt 
\lower3.06pt\hbox{{$\scriptstyle \sim$}}}}         
\newcommand{\ls}{\mathrel{\raise1.16pt\hbox{$<$}\kern-7.0pt 
\lower3.06pt\hbox{{$\scriptstyle \sim$}}}}         

\newcommand{\hMpc}{\;h^{-1}{\rm Mpc}}
\newcommand{\kMpc}{\;h\,{\rm Mpc}^{-1}}

\input epsf
\def\japfigA#1{
\epsfxsize=8.4cm
\epsfbox[101 15 434 790]{bphfig#1.ps}
}
\def\japfigB#1{
\epsfxsize=8.4cm
\epsfbox[25 100 485 610]{bphfig#1.ps} 
}
\def\japfigC#1{
\epsfxsize=8.4cm
\epsfbox[15 120 520 665]{bphfig#1.ps} 
}

\begin{document}

\maketitle

\begin{abstract}
It has been proposed that the cosmological constant $\Lambda$ might
be measured from geometric effects on large-scale structure. A positive
vacuum density leads to correlation-function contours which are squashed in
the radial direction when calculated assuming a matter-dominated model.
We show that this effect will be somewhat harder to detect than
previous calculations have suggested: the squashing factor is
likely to be $<1.3$, given realistic constraints 
on the matter contribution to $\Omega$.
Moreover, the geometrical distortion risks being confused with the
redshift-space distortions caused by the peculiar velocities
associated with the growth of galaxy clustering. These depend on
the density and bias parameters via the combination
$\beta\equiv \Omega^{0.6}/b$, and we show that the main practical
effect of a geometrical flattening factor $F$ is to simulate
gravitational instability with $\beta_{\rm eff}\simeq 0.5(F-1)$.
Nevertheless, with datasets of sufficient size it is possible
to distinguish the two effects. We discuss in detail how this
should be done, and give a maximum-likelihood method for
extracting $\Lambda$ and $\beta$ from anisotropic
power-spectrum data. New-generation redshift surveys of
galaxies and quasars are potentially capable of detecting a
non-zero vacuum density, if it exists at a cosmologically
interesting level.
\end{abstract}

\section{INTRODUCTION}

In recent years there has been a resurgence of interest in the cosmological
constant $\Lambda$ as a possible way of evading several cosmological problems
(see Carroll \etal\ 1992 for a review).
Although long popular with theorists, the
Einstein-de Sitter $\om = 1$ model (hereafter denoted EdS) now
seems increasingly untenable owing to its short expansion timescale.
An open model with low matter density parameter 
$\om$ and no cosmological constant allows a
younger universe, but is more difficult to reconcile with inflation.
Although some workers have suggested that bubble nucleation within
inflation may be capable of yielding an open universe
(e.g. Bucher, Goldhaber \& Turok 1995), a more
common alternative to the EdS model is to retain a flat universe, $k=0$,  through
$\Omega \equiv \om+\ov=1$. A high value of $\ov\simeq 0.8$ would then be
indicated by arguments for $\om\simeq 0.2$ from cluster $M/L$ 
values and large-scale structure measurements (Efstathiou, Sutherland \& Maddox 1990).
In any case, it must be remembered that there is no known reason for
the vacuum density to vanish (Weinberg 1989), and so it is
profoundly important for physics to test whether $\lam$ is non-zero.

It has been suggested by Alcock \& Paczy\'{n}ski (1979) that it may be possible to
detect the presence of $\Lambda$ by a geometric test, measuring the effect of
deviations from the assumed EdS geometry on large scale
structure. The assumption of an incorrect geometry can lead to an effective
squashing of space along the line of sight -- causing an anisotropy in the
inferred density field which could be detected from galaxy (or quasar)
clustering statistics. This has the crucial advantage over 
other tests such as number counts
of being independent of galaxy or quasar evolution. Phillipps (1994) considered
the possibility of analysing a quasar survey using the orientation of pairs and
claimed that the effect should be readily detectable. However, the only 
cosmological constant model discussed by Phillipps is de Sitter space:
zero mass density and $\ov = 1$. We show that if the matter density
parameter is even modestly greater than zero then the geometric effect is much
reduced and also does not continue to increase with redshift above $z \simeq 1$.

Redshift-space distortions caused by peculiar velocities of galaxies also lead
to anisotropic structure. Large-scale infall squashes overdensities along
the line of sight in redshift space, which can mimic the geometric squashing
caused by $\Lambda$. In the linear regime this distortion has a simple effect
in Fourier space in the `distant observer' approximation (see Kaiser 1987) and
is characterised by the parameter $\beta = \om^{0.6}/b$ (where $b$ is the
bias parameter). For parameters of interest, redshift distortion is not
negligible in comparison with the geometric distortion.
The situation is complicated further on small scales, where
virialized clusters appear elongated 
along the line of sight -- the so-called `fingers of God'.

We therefore consider in some detail the clustering anisotropies which arise
in the presence of all three effects: geometric flattening, $\beta$-distortion
and fingers of God. We use a power spectrum analysis, since the 
modelling of redshift-space  distortions is simpler
in Fourier space than it is in real space.
Although the Kaiser and squashing effects are similar, the functional forms
of the anisotropies in the power spectrum differ, and $\Lambda$ and $\beta$ can
be distinguished in principle.
Sections 2 \& 3 plus Appendix A give the basics of the effect.
In Section 4, we present
a maximum-likelihood technique for measuring both $\lam$ and
$\beta$ from these anisotropies in Fourier space.
Finally, in Section 5, we give some assessment of the likely practical
constraints on $\Lambda$ that may be expected from next-generation
redshift surveys, such as the Anglo-Australian 2-degree field
survey and the Sloan Digital Sky Survey.

\section{Cosmological Models and Clustering Anisotropy}

To measure the cosmological constant, we exploit the fact that we cannot
measure comoving distances directly, but use redshift as a distance indicator.
An object which is spherical in comoving real space $\bf r$ will 
only appear spherical to
an observer (who measures redshifts) if the correct geometry is assumed -- \ie\
the correct $r(z)$ relation is used.

We write the Robertson-Walker metric as:
\be
{\rm d}s^{2} = c^{2}{\rm d}t^{2} - R(t)^{2}\left[{\rm d}r^{2}
+ S_{k}^{2}(r)\left({\rm d}\theta^{2} + \sin^{2}\!\!\theta{\rm d}\phi^{2}
\right)\right].
\ee
Here $R_{0}r$ is the comoving geodesic distance, $R(t)$ is the cosmic
scale factor ($= R_{0}$ now) and :
\bd 
S_{k}(r) = \left\{ \begin{array}{lll}
			\sin r & k = 1 & \mbox{closed} \\
			r & k = 0 & \mbox{flat} \\
			\sinh r & k = -1 & \mbox{open}
			\end{array}
	\right. 
\ed
The square of the comoving distance between two objects at $z \pm
\Delta z/2$ separated by $\Delta\theta$ is:
\bd
\Delta \ell^{2} = R_{0}^{2}S_{k}(r)^{2}\Delta\theta^{2} +
R_{0}^{2}\left(\frac{{\rm d}r}{{\rm d}z}\right)^{2}\Delta z^{2}
\ed
For a universe with matter contribution $\Omega_{m}$ and cosmological
constant contribution $\ov$:
\bd
\longeq{
R_{0}\frac{dr}{dz}(z) \equiv A_{\rm t}(z)
}
{
= \frac{c}{H_{0}}\frac{1}{\sqrt{(1-\Omega)(1+z)^{2}+\ov + \om(1+z)^{3}}} 
}
\ed
\bd
\longeq{
R_{0}S_{k}(r)\equiv B_{\rm t}(z) = \frac{c}{H_{0}} |1-\Omega|^{-1/2} \times
}
{
S_k \left[\int_{0}^{z}\!\!\frac{|1-\Omega|^{1/2} \; dz\pr}{\sqrt{(1-\Omega)(1+z\pr)^{2}+\ov +
\om(1+z\pr)^{3}}}\right]
}
\ed
For flat models, favoured by inflation, $\Omega = \om+\ov = 1$, so:

\bd
R_{0}\frac{dr}{dz}(z) = \frac{c}{H_{0}}\frac{1}{\sqrt{(1-\om) + \om(1+z)^{3}}}
\ed

\be
R_{0}r(z) = \!\!\frac{c}{H_{0}}\int_{0}^{z}\!\!\frac{dz\pr}{\sqrt{(1-\om) +
\om(1+z\pr)^{3}}}
\ee
Even with this simplification, the integral for $R_0r(z)$ must be carried
out numerically unless $\ov=0$ or $\om=0$.

There are three coordinate systems to consider here. We denote by the
subscript t the true geometry, where $z$ represents the redshift from pure
Hubble expansion at the coordinate distance $r$.
We denote by s the coordinate system where the geometry is correct, but the
redshift is used as a distance indicator. This incorporates redshift distortion
from peculiar velocities.
Finally, we denote by the subscript a the assumed geometry. This will always be
an EdS model in the present paper, and quoted flattening factors are given relative to this model,
for which we have
\bd
R_{0}\frac{dr}{dz}(z) = \frac{c}{H_{0}}\frac{1}{(1+z)^{3/2}} \equiv A_{\rm
a}(z)
\ed
\bd
R_{0}r(z) = 2\frac{c}{H_{0}}\left(1-\frac{1}{\sqrt{1+z}}\right) \equiv B_{\rm
a}(z)
\ed

\noindent Taking EdS as default, we define geometric squashing factors:

\bd
\fpl(z) \equiv \frac{A_{\rm t}}{A_{\rm a}}
\ed

\bd
\fpp(z) \equiv \frac{B_{\rm t}}{B_{\rm a}}
\ed

\bd
F(z) \equiv \frac{\fpl}{\fpp} = 1 + \frac{1}{4}(1-\om+2\ov)\, z
+ O(z^{2})
\ed

\noindent $\fpp,\fpl$ and $F$ are all unity for EdS.
The flattening factor $F$ is defined so that $F>1$ means that
objects would appear flattened along the line of sight in the
assumed geometry. The first-order redshift dependence
(obtained with the aid of {\it Mathematica\/}) tells us that
what is really measured via the flattening is not $\ov$ but
$\ov-\om/2$. This is inevitable, since a general argument
(e.g. section 14.6 of Weinberg 1972) shows that the lowest-order
corrections to the distance-redshift relation
depend only on $q_0 = \om/2 - \ov$.  The accuracy with which one can measure $\ov$
from low-redshift data is then limited by (amongst other things) one's
knowledge of $\Omega_m$, which is currently rather poorly known.
In principle, measuring the evolution at several redshifts
out to $z>1$ would allow both $\om$ and $\ov$ to be determined separately, but this
will be harder. 
It is also interesting to compare the $\om$ dependence
in the case of open and flat models: $F\simeq 1+(1-\om)z/4$
with $\ov=0$ but the three-times stronger $F\simeq 1+3(1-\om)z/4$ for $k=0$.

\begin{figure}
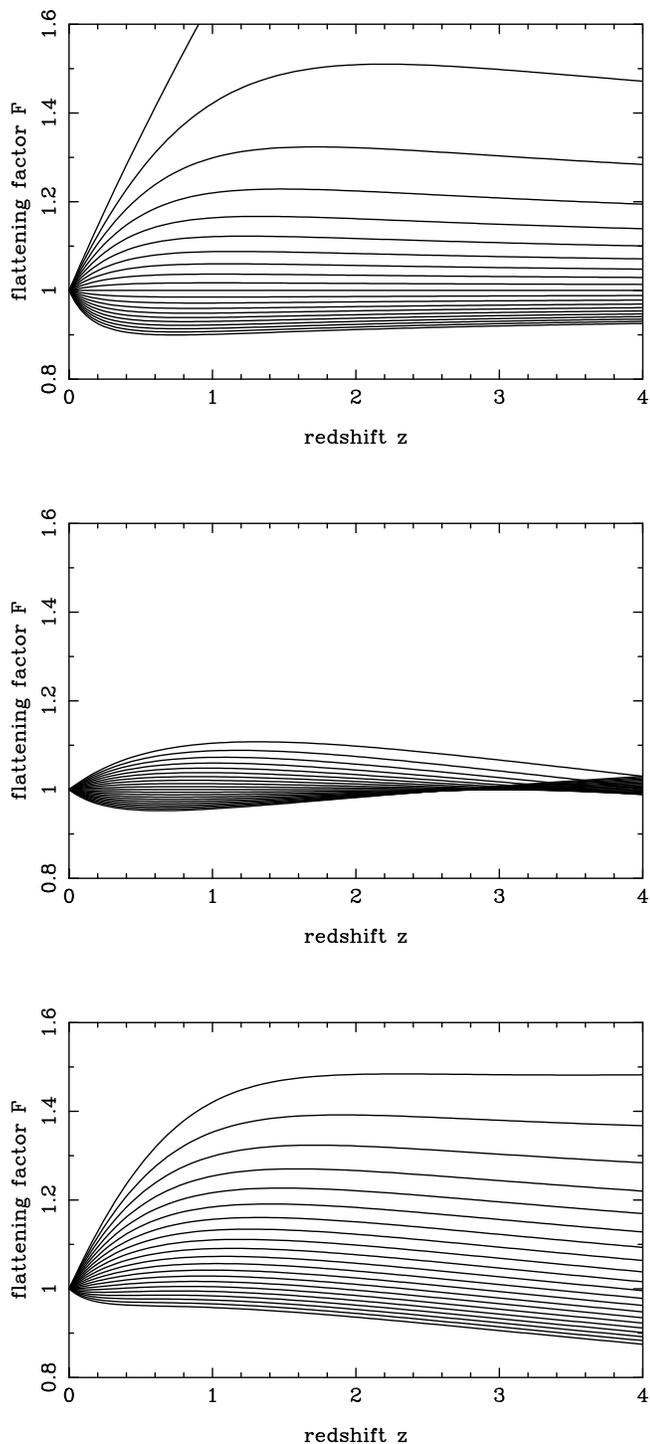

\japfigA{1}
\caption{(a) Flattening factor $F$ as a function of redshift for flat models
($\om + \ov = 1$). The curves range from $\ov = 1$ (top curve) to $\ov = -1$ in
steps of $0.1$.
(b) $F$ against redshift for models with a fixed mass density $\om = 0.2$ and a
cosmological constant. $\ov$ varies as for (a).
(c) $F$ against redshift for models with no cosmological constant. $\om$ varies
from 0 (top) to 2, again in steps of 0.1.}
\end{figure}

It can be seen from Fig. 1a and Fig. 1b that the squashing effect of $\Lambda$
(characterised by F) increases rapidly with redshift and then saturates at
redshift of $z \simeq 1$ for all but pure de Sitter space ($\om=0$). The fact
that $F$ does not increase without limit if $\om \neq 0$ is easily understood.
At high redshifts matter dominates over vacuum energy because of their
equations of state, and

\be
\frac{\om(z)}{\ov(z)} \propto (1+z)^{3}.
\ee

\noindent Hence at high $z$, the cosmological constant becomes insignificant.
The saturation is quite severe: $F \leq 1.3$ for $\om \geq 0.2$ (the limit of
most current models). Compare this with the de Sitter model which reaches $F =
2.2$ at $z = 2$.
There is therefore unlikely to be any gain by going to very high redshifts: a
galaxy survey with a large number of galaxies may well be preferable to a
deeper but much smaller quasar survey.
The calculations can be repeated for open models with no cosmological constant,
but the effects are much smaller (see Fig. 1c). Hence a large value of $F$
indicates a cosmological constant, and would be a valuable robust measure of
$\lam$, if it were not subject to confusion with other effects.

To utilise the squashing as a diagnostic for the geometry, it is necessary to
have something of known shape. A convenient such object is the power spectrum
of galaxy (or quasar) clustering, whose contours in {\bf k}-space must be
spherical if the cosmological principle holds
(although a survey which covered a significant range of redshifts
would require  power to be measured separately in a number of redshift 
bins to avoid anisotropy introduced by evolution).
Denoting the real comoving wavenumber by {\bf k}, and splitting it into a
component $k_{\parallel}$ along the line-of sight and a component 
${\bf}k_{\perp}$
perpendicular to it, the true comoving real space power spectrum will be a
function only of $|k|$:

\bd
P_{\rm t}(k_{\parallel},{\bf k}_{\perp}) = P_{\rm t}(k).
\ed

\noindent 
If we assume the wrong geometry, we measure power at the wrong
wavelengths. Ignoring redshift distortions for the time being:

\be
P_{\rm a}(k_{\parallel},{\bf k}_{\perp}) = \frac{1}{\fpp^{3+n}F}P_{\rm
t}(k)\left[1+\mu_{\rm a}^{2}\left(\frac{1}{F^{2}}
-1\right)\right]^{\frac{n}{2}},
\ee
where $n$ is the local spectral index of the power spectrum and
$\mu_{a} =$ cosine of wavevector to line of sight. This
expression is derived in Appendix A, but is immediately
reasonable: $P(\mu_a=0)/P(\mu_a=1)=F^n$, as expected for
squashing by a factor $F$. 
This equation implies that there is
no sensitivity to geometry for an $n=0$ spectrum;
however, this ceases to be true in the presence of
peculiar velocities, which also give rise to
power-spectrum anisotropies.

\section{Redshift Distortions}

\subsection{Linear Modes}

In models where structure forms via gravitational instability, density
fluctuations inevitably induce peculiar velocities, which affect the mapping to
redshift space where the radial coordinate is a total velocity.
In linear theory, density waves are boosted along the line of  sight in redshift
space (Kaiser 1987).
We assume that the survey subtends a small angle in the sky and lines of sight
can be treated as parallel, so that the simple Kaiser  formula remains valid (see
Heavens \& Taylor 1995 and Zaroubi \& Hoffman 1996 for more general analyses) -- \ie\ even with the correct 
assumed geometry, there is still anisotropy in redshift space:

\be
P_{\rm s}(k,\mu)
= P_{\rm t}(k)[1+\beta\mu^{2}]^{2}
\ee

\noindent where $P_{\rm t}$ is the isotropic real-space power
spectrum and $P_{\rm s}$ is the power spectrum in the correct geometry.

This boosts the power spectrum along the line of sight as does the $\Lambda$
effect, but it is clear from formulae (15) and (16) that the $\mu$-dependence is
different.
Gravitational instability generates a characteristic ratio between
the $\mu^2$ and $\mu^4$ components, which differs in general from that
resulting from geometrical distortion.
Hence the two effects are distinguishable in 
principle, given data of sufficiently high signal-to-noise (see Fig. 2).

\begin{figure}
\japfigB{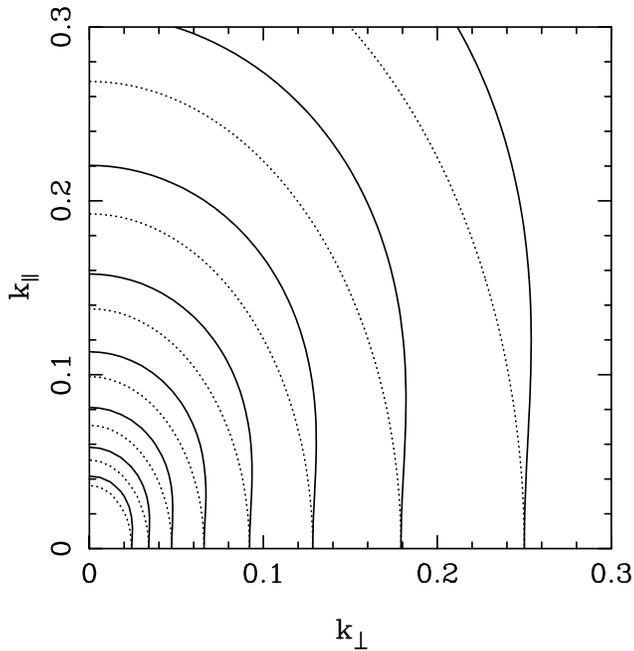}
\caption{ Contours of the power spectrum in the $k_{\parallel}$ and $k_{\perp}$
plane, assuming a power-law index $n=-1.5$ for the true power spectrum.
The contour interval is $\Delta {\ln P} = {1/2}$. The models are linear (Kaiser) redshift distortion only with $\beta = 0.5$ (full contours) and geometric squashing effect only, with $F=1.5$ (dotted contours). }
\end{figure}

\subsection{Fingers of God}

Additionally, we consider the nonlinear finger-of-God effect caused by velocity
dispersions in virialized clusters. 
A simple model for the effects of such velocities on the power
spectrum was introduced by Peacock (1992; see also Peacock \& Dodds 1994).
This consists of multiplying the linear-theory Kaiser distortion by a 
term which treats the radial distortion as a convolution with
an incoherent velocity component, leading to damping of power at large
values of $k_\parallel$:
\be
P_{\rm s}(k,\mu) = 
P_{\rm t}(k)[1+\beta\mu^{2}]^{2}\; D[k\mu\sigma_{\rm p}].
\ee
If the dispersion is taken to be Gaussian, the damping term is
\be
D[k\mu\sigma_{\rm p}]=\exp[-k^2\mu^2\sigma_{\rm p}^2/2].
\ee
Here $\sigma_{\rm p}$ is the line-of-sight {\it pairwise\/} dispersion in relative
galaxy velocities caused by the the incoherent dispersion
($\sigma_{\rm p}=\sqrt{2}\sigma_v$). This is usually quoted
in velocity units, but is implicitly divided by 100 to obtain
a scale-length in $h^{-1}\rm Mpc$ when used in power spectra\footnote{Hubble constant 
$H_0 = 100\,h\;$kms$^{-1}$Mpc$^{-1}$}.
Note that it is not obvious that this parameter corresponds exactly 
to the pairwise velocity dispersion measured using other methods, although
it should be close.
In reality, the pairwise velocity distribution is better modelled by an
exponential (Davis \& Peebles 1983), which leads to a Lorentzian factor in Fourier space:
\be
D[k\mu\sigma_{\rm p}] =
\frac{1}{1+\left(k\mu\sigma_{\rm
p}\right)^{2}/2}.
\ee
In practice, there is very little difference between these models until
the damping becomes a factor $\gs 2$ -- see Fig. 3.
According to a comparison with $N$-body simulations by Cole, Fisher \& Weinberg 
(1994) and (1995) (hereafter CFW94 and CFW95), this simplified analytical
model appears to work well up to $k\sigma_{\rm p}$
of order unity. We note that
CFW95 in fact use a damping factor of a Lorentzian squared, which is not equivalent
to an exponential pairwise distribution (it actually corresponds to an
exponential one-particle distribution); however, this makes very little
difference to the answers.

\begin{figure}
\japfigB{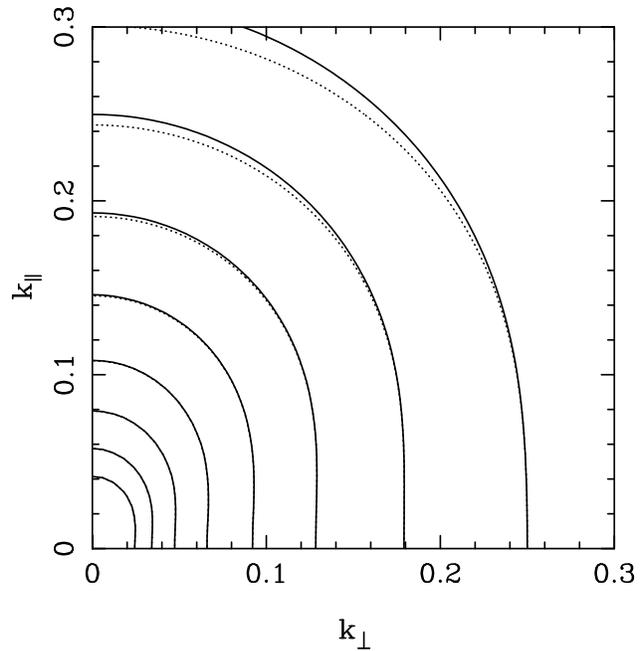}
\caption{ Contours for both linear and nonlinear redshift distortions with
$\beta = 0.5$, and $\sigma_{\rm p} = 350\; {\rm kms}^{-1}$ (again for $n = -1.5$). The solid contours are for
exponential small scale velocity distribution, the dotted Gaussian.
The contour interval is $\Delta {\ln P} = {1/2}$.
The stretching of contours of the correlation function along the line of sight
leads to a reduction in line of sight power on small scales (large k) -- see
also Cole, Fisher \& Weinberg (1994).}
\end{figure}

\subsection{The general case}

Combining the effects of geometry, linear peculiar velocities and
fingers of God, it is shown in Appendix A that the overall effect is

\bean
P_{\rm a}(k,\mu)
&=& \frac{1}{\fpp^{3+n}F}P_{\rm
t}(k)\left[1+\mu^{2}\left(\frac{1}{F^{2}}-1\right)\right]^{\frac{n-4}{2}}
\nonumber \\
 & \times& \left[1+\mu^{2}\left(\frac{\beta+1}{F^{2}}-1\right)\right]^{2}
D[k\mu\sigma_{\rm p}\pr],
\eean

\noindent where $\sigma_{\rm p}\pr \equiv \sigma_{\rm p}/\fpl$, and we have modified
$\mu$ to account for the assumed geometry. 
If $\beta=0$, and $F$ is close to unity, then the anisotropy
resembles a pure redshift-space distortion, with an apparent value of $\beta$:
\bd
\beta_{\rm a} \simeq -{n\over2}\,(F-1).
\ed
If the true $\beta$ is small but non-zero, then the true distortion
adds linearly to the geometrical signature;
for large $\beta$, things
are more complex, since $F$ also modifies the effect of $\beta$.
Since the effective $n\simeq -1$ on the scales of interest where $\beta$
can be measured, we see that any experiment which hopes to detect
$\Lambda$ through $F\simeq 1.3$ will need to be able to measure $\beta_{\rm a}$
to a precision of rather better than $\pm 0.1$. 
Fig. 4 shows that the differences
between the $F$ and $\beta$ effects are subtle;
a large survey with good statistical signal-to-noise will be needed
if the two effects are to be measured separately.
Furthermore, because the effects of interest are small, careful control of
systematics will be required. For example, one will not wish to assume
the precise correctness of the model (17) for the redshift-space
distortions, and empirical relations from numerical simulations will
be more appropriate. However, for the present the simple model
suffices to indicate how hard we will have to work.  It also illustrates how 
well one needs to know the intrinsic shape of the objects whose flattening is
being exploited.  The scatter in the shapes of cosmic objects such as voids (cf.
Ryden 1995) will preclude their use for this purpose.

\begin{figure}
\japfigB{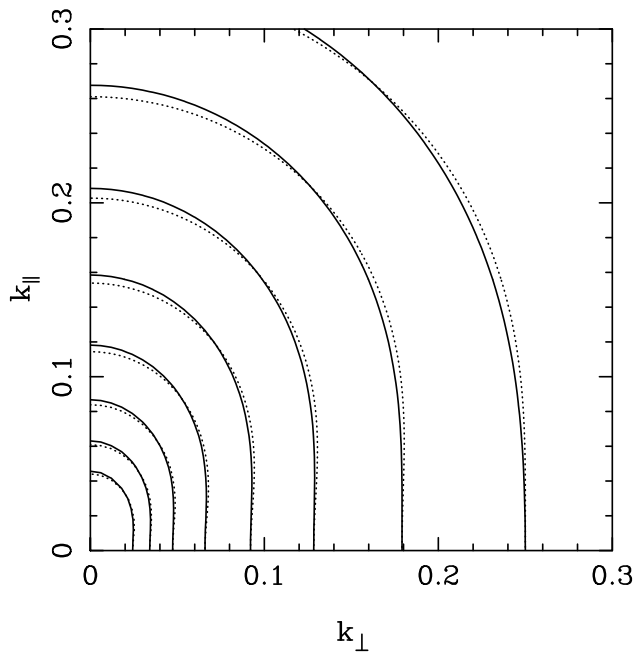}
\caption{Solid contours for a model with $F=1.1$, $\beta = 0.5$ and
$\sigma_{\rm p} = 350 \;{\rm kms}^{-1}$. Dotted contours are for the best fit to this model
with redshift distortions only ($F=1$).}
\end{figure}

\subsection{Geometry from evolution of $\beta$}

Although we have shown that $F$ and $\beta$ are degenerate
to some extent at a single redshift, they are expected
to evolve differently. This in itself does not help to
detect $\Lambda$, because the evolution of bias is unknown:
\bd
\beta(z) = {\Omega_m^{0.6}(z)\over b(z)}.
\ed
Nusser \& Davis (1994) have argued that $b(z)$
is calculable provided galaxy numbers are preserved; however,
we feel it is better not to assume that merging is negligible
and to treat the evolution of bias empirically.
This argues that we must seek a signature of
$\Lambda$ which takes the empirical $\beta(z)$ without
seeking to assume what fraction of the evolution is
due to changes in $\Omega_m$ with redshift. The following sections
of this paper are devoted to seeing what can be done
by looking at the detailed $\mu$-dependence of the
clustering anisotropies, in order to break the
lowest-order degeneracy.

However, further information on $b(z)$ can be obtained from the
evolution of the amplitude of the power spectrum with redshift,
as discussed by Matsubara \& Suto (1996).
In the linear regime, the true galaxy power spectrum evolves as
\bd
P_{\rm t}(z)=b^2(z)P_{\rm mass}(z)={b^2(z)\over b^2(0)}\,{\cal D}^2(z)\,P_{\rm t}(0),
\ed
where ${\cal D}(z)$ is the linear-theory growth law for density perturbations.
Unfortunately, this evolution is not directly measurable:
even looking at real-space clustering, there are
geometrical factors which alter the inferred clustering (cf. the $\mu=0$ limit
of equation [20]); what is observable is thus the apparent growth factor
\bd
{\cal D}_{\rm a}(z)={b(z)\over b(0)}\,{ {\cal D}(z) \over [f_\perp^{3+n}(z)F(z)]^{1/2}}.
\ed

The other observable is the apparent redshift-space distortion as a function
of redshift, which we choose to describe by an apparent value of $\beta$.
If the true $\beta$ is of order unity, then the geometrical and velocity
distortions interfere in a complex manner. Suppose we deduce an apparent
$\beta$ by fitting $P(\mu)/P(\mu=0)$ by a velocity-only distortion;
to first order in $(F-1)$, this gives an apparent value of $\beta$
\bd
\beta_{\rm a}(\mu)=\beta - \left[ 2\beta(1-\mu^2) + {n\over 2}(1+\beta\mu^2)\right]\,(F-1).
\ed
There are thus many possible choices of $\beta_{\rm a}$; the simplest
procedure is probably to average these choices  over $\mu$, yielding
the redshift-dependent observable
\bd
\beta_{\rm a}(z)=\beta(z) - {8\beta(z) +[3+\beta(z)]n \over 6}\,(F(z)-1).
\ed
For a given hypothetical $(\om, \ov)$ pair, the equations for $\beta_{\rm a}(z)$
and ${\cal D}_{\rm a}(z)$ both yield $b(z)/b(0)$; requiring these curves to match
at all $z$ can in principle allow both density parameters to be determined.

It is interesting to explore this possibility in a little more detail
by looking at the first-order change with redshift of these relations.
The initial change with redshift of the observables are respectively
\bd
{d{\cal D}_{\rm a}\over dz}=-\om^{0.6}-{n+4\over 8}\,[1-\om+2\ov] + {d\ln b\over d z}
\ed
and 
\bd
{d\beta_{\rm a}\over dz}=\left[{4\beta\over 15}-{n\over 24}(3+\beta)\right]\,(1-\om+2\ov) - \beta\,{d\ln b\over d z}.
\ed
The second of these depends just on $q_0$, but the first involves
a different combination of $\om$ and $\ov$.
If we eliminate the unknown bias evolution $d\ln b / d z$ from the above equations, we get
\bd
\om^{0.6}=-{d{\cal D}_{\rm a}\over dz} - {d\ln\beta_{\rm a}\over dz} -
\left[{7\over 30} + {n\over 6} + {n\over 8\beta_{\rm a}}\right]\, (1-2q_0).
\ed
For plausibly realistic values $n\simeq -1$, $\beta\simeq 0.5$,
the coefficient of $(1-2q_0)$ is very small,  giving a diagnostic for $\om$ alone.
Studies of clustering evolution and anisotropy out to $z\simeq 0.5$ may
thus principally determine $\om$, rather than $\ov$. In order to pin down
the cosmological constant, it will be necessary either (a) to work at
$z\gs 1$; (b) to work at large scales where $n\simeq 0$; or (c) to
look in more detail at the angular dependence of the anisotropies.
We now return to the last of these options.

\section{Maximum Likelihood Fitting}

Given the model (20) for the combined geometrical plus 
redshift-space distortions, the standard way of proceeding would be to
perform a maximum-likelihood fit on a dataset to obtain
the required parameters $\beta$,$F$,$\sigma_{\rm p}\pr =
\sigma_{\rm p}/\fpl$ and a parameterized form for $P_{\rm t}(k)$ simultaneously.
Smaller uncertainties in the most interesting parameters such as $\beta$
and $F$ may be obtained if we use a priori knowledge of e.g. the
shape of $P_t$, but
in order to make a robust test for $\Lambda$, it is safer to obtain
all the information we require from the same survey.

To apply maximum-likelihood methods, we need to know the 
probability distribution for the power.
The central limit theorem implies that the observed power 
averaged over many independent modes
will have a Gaussian distribution (even though single modes will
be Rayleigh distributed for a Gaussian random field -- see Feldman \etal\ 1994).
Similarly, $\ln P$ would be expected to have a
Gaussian distribution with variance $\sigma^{2} = 1/N$ where $N$ is the number
of independent modes averaged over:

\be
\longeq{
f(\ln \! P_{\rm a}) \; {\rm d}\ln \! P_{\rm a} =
\left(\frac{N}{2\pi}\right)^{\frac{1}{2}} \times 
}
{
\exp\left[-\frac{N}{2}(\ln \! P_{\rm
a}-\ln \! P_{\rm a}^{\prime})^{2}\right]{\rm d}\ln \! P_{\rm a},
}
\ee
where $P_{\rm a}$ is the observed power as a function of $k$ and $\mu$,
$P_{\rm a}\pr[k,\mu;\lam,\beta,\sigma_{\rm p}\pr,P_{\rm t}(k)]$ is the
corresponding mean power of the model.
In both cases, the powers are the {\it total\/} power, including shot noise.
The likelihood estimator is then

\be
\longeq{
\ln{\cal L}[\lam , \beta , \sigma_{\rm p}\pr,P_{\rm t}(k)] = {\rm const.} - 
}
{
\frac{1}{2}\sum_{\rm modes}(\ln \! P_{\rm a} -\ln \! P_{\rm a}
^{\prime})^{2} ;
}
\ee
in the continuum approximation, this becomes:

\be
\longeq{
\ln{\cal L} = {\rm const.} - 
}
{
\rho\int_{0}^{k_{\rm max}}\!\!\pi\, k^{2} \; dk
\int_{0}^{1}(\ln \! P-\ln \! P^{\prime})^{2} \; d{\mu},
}
\ee
where $\rho$ is the density of states in {\bf k}-space.

For a non-uniform selection function, the effective density of states is
(Appendix B)
\be
\rho_{\rm eff} = {[\int w\,\bar n\; d^3r]^2\over (2\pi)^3 
\int w^2\bar n^2\; d^3 r}.
\ee
where $\bar n(r)$ is the mean density of objects, and $w(r)$ is the
optimal weighting function of Feldman et al. (1994) for measuring power: 
$ w=(1+\bar n P)^{-1}$.
The integral is restricted to a maximum value of $k = k_{\rm max}$ to prevent
excessive nonlinearity, in particular to avoid wavenumbers where the ansatz for
the fingers of God may not be a good approximation. CFW95 showed that this
limit corresponds to $k\sigma_{\rm p}\pr \ls 1.6$ (10 per cent error in
the derived $\beta$). 
The precise maximum wavenumber allowed by this criterion depends on the
assumed $\sigma_{\rm p}$. Taking a conservative view of this, we will
generally set $k_{\rm max} = 0.35\kMpc$ as a safe limit
within which $F$ is not significantly affected by systematics in
the inferred $\beta$.
The sum/integral is over a hemisphere in $k$-space. The modes in the other
hemisphere are identical: $P({\bf k}) = P(-{\bf k})$ because the density field
is real.

The procedure in practice is to maximise ${\cal L}$ to find the best fitting
parameters $\lam$, $\beta$, $\sigma_{\rm p}\pr$ and $P_{\rm t}(k)$. We assume a
shape for $P_{\rm t}(k)$ and fit only for the amplitude, although the shape
could be parameterized for a more extensive analysis. In this paper we are
interested principally in the errors on the parameters, to work out whether any
given survey will be able to distinguish between cosmological models of
interest.

\subsection{Covariances and Correlation matrix}

We now wish to study the form of the likelihood contours that
describe the joint distribution of the interesting parameters
in this problem. This question can be studied in the absence of data
because of the simple form of the likelihood function, which
is a sum of squared differences in $\ln P$, multiplied by the
density of states in $k$ space. For a given observed power,
the likelihood thus scales simply with the volume of
the survey; the likelihood contours are of a fixed form,
which are merely re-labelled as the number of modes increases.
The covariance matrix $C_{ij}$ for the parameters $a_i$ is thus of a universal
shape which can be scaled once the absolute value of the error on
one parameter is known
(at least for a fixed ratio of true power to
shot power).

Even the effect of shot noise is rather weak. It only enters
at all because the error in power scales with the power:
$\sigma(P)$ thus changes with $\mu$ in the presence of
redshift-space anisotropies, provided shot noise is negligible.
Conversely, if shot-noise dominates, $\sigma(P)$ is a constant.
In one limit, therefore, the likelihood measures the sum of
$[\Delta\ln P]^2$, whereas in the other it is $[\Delta P]^2$.
In practice, the difference seems to be small, as shown in Figs 5 \& 6.
The optimally-weighted QDOT dataset has $P_{\rm shot}=P_{\rm true}$
at $k=0.05\kMpc$ and $P_{\rm shot}=10 P_{\rm true}$ at $k=0.2 \kMpc$.
Future large-scale surveys are likely to have negligible shot noise
on all quasi-linear scales.

To produce likelihood contours requires some data for the
`observed' power. In what follows, we simply take this to be
the expected power in the model under consideration. A more exact
procedure would be to produce a realization of the model, with
the power for each mode exponentially distributed about the
expectation value. The maximum-likelihood value of the parameters $a_i$
obtained by fitting to such a fictitious dataset would then differ
from the input ones, within the `error bars' produced by the
likelihood analysis. However, these errors are the same
whether we input the expectation spectrum as data, or use
a realization:
\be
\longeqtwo{
\Delta\ln{\cal L} =  -{1\over 2}\sum (\ln P_{\rm obs} - \ln P_{\rm mod})^2
}
{\Rightarrow
C^{-1}_{ij}=-\left.{ \partial^2 \ln{\cal L} \over \partial 
a_i \partial a_j}\right|_{ML} \simeq
\sum { \partial \ln P_{\rm mod} \over \partial a_i} { \partial 
\ln P_{\rm mod} \over \partial a_j},
}
\ee
independent of the data.   Note that this expression is only approximately 
true, since it assumes that the term $\sum (\ln P_{\rm obs} - \ln P_{\rm mod}) 
\partial^2P_{\rm mod}/\partial a_i \partial a_j$ is negligible.  The maximum
likelihood solution ensures that a somewhat different weighting of the data 
is zero:  $\sum (\ln P_{\rm obs} - \ln P_{\rm mod}) 
\partial P_{\rm mod}/\partial a_i = 0$.  In what follows in this section, we 
neglect the additional term;  the numerical results presented do not make this
approximation.   Using the expectation power spectrum as data 
therefore gives approximately the right error contours on the parameters, but
centred on the input data. The constraints that can be
set from a given real dataset would then depend on the actual 
maximum-likelihood values, which we would expect to be displaced 
from the true value
by about the `one-sigma' error calculated here.

Because the model power spectrum studied here is of a rather simple form,
the covariance matrix can be obtained analytically in many cases, at least
with the aid of a symbolic manipulator such as {\it Mathematica\/}. The
general expressions are sufficiently messy that they are not worth
reproducing, but there are a few exact results that are simple enough to be 
useful. First recall
that the power to be fitted is the sum of a true power $P_T$ plus 
shot power $P_S$;
one might therefore approximate the covariance matrix as follows, simply
interpolating between the scaling of the results for zero shot noise and for
dominant shot noise: 
\be
C_{ij}={2\over N}\left( C^A_{ij} + {P_S\over P_T} C^B_{ij} \right),
\ee
where $N$ is the total number of modes being considered (over a complete sphere
in $k$ space). The covariance matrices for the shot-free and shot-dominated
cases ($C^A$ and $C^B$ respectively) depend on the number of parameters
being fitted for. Perhaps the most optimistic case would seem to be
the two-parameter model where only $\beta$ and $F$ are allowed to vary,
assuming that both the damping $\sigma_p$ and the amplitude of
the power spectrum are known. The latter assumption is probably unreasonable,
since the main manifestation of a non-zero $\beta$ or non-unity $F$ is
to change the mean level of power; the real signature we are interested
in is the power anisotropy. Within the likelihood methodology, this
demands that we allow the amplitude of the spectrum to float as a third
parameter which is integrated over when we consider the distribution of the 
interesting parameters $\beta$ and $F$. 

\begin{figure}
\japfigC{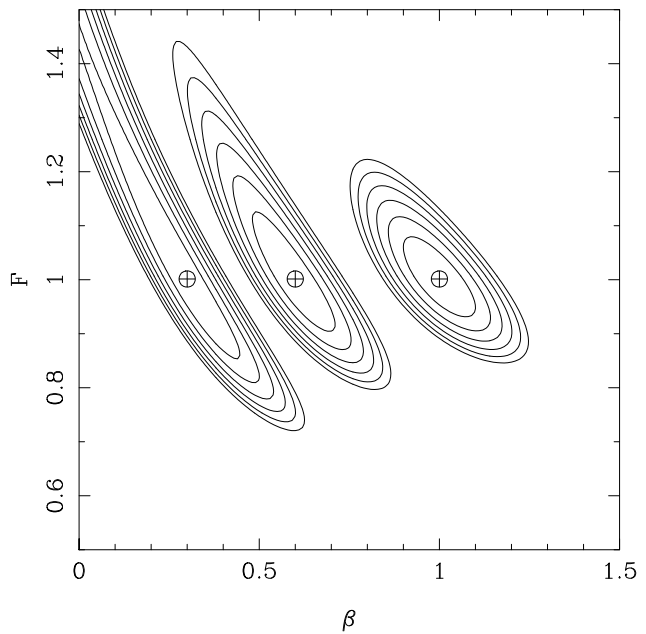}
\caption{Expected contours of likelihood in the $F-\beta$ plane,
for the case of a survey with negligible shot noise, for three values of 
$\beta$.
The scaling is set so that $\sigma(\beta)=0.1$ for $\beta=1$.
At each $(\beta,F)$ point, the maximum-likelihood value of
the power-spectrum amplitude has been chosen.
The contour interval is $\Delta{\ln {\cal L}} = 1/2$. The position of the true 
values of the parameters is marked by the cross; 
contours are shown for  $\beta = 0.3$, $0.6$ and $1.0$.
}
\end{figure}

\begin{figure}
\japfigC{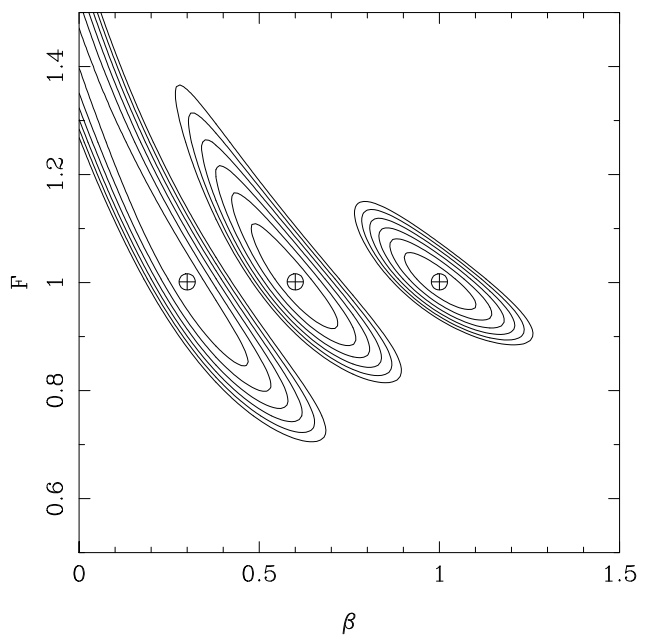}
\caption{
Expected contours of likelihood in the $F-\beta$ plane,
as for Fig. 5 except now
for the case of a survey dominated by shot noise.
}
\end{figure}

For this model, the covariance matrices in the shot-free and
shot-dominated limits
can be obtained analytically in the
limit of small $\beta$, where both have the same form:
\be
C^A_{\beta\beta}=C^B_{\beta\beta}={11025\, n^2\over 256\,(-4+n)^2}\;\beta^{-2}.
\ee
As we saw earlier when considering the quadrupole term, in this limit
\be
C_{FF}={4\over n^2}\,C_{\beta\beta},
\ee
so that $\beta$ can be obtained with greater accuracy than $F$ if 
$n \simeq -1$.
If only $\beta$ and $F$ had been allowed to vary, both results
would have been a factor $16/25$ smaller -- reflecting the
spurious accuracy that arises if the effects on the normalization
of the power spectrum are ignored.
As $\beta$ increases, $\beta^2 C$ changes approximately linearly
with $\beta$, and it will suffice to quote the results at $\beta=1$.
For the shot-free case:
\be
C^A_{\beta\beta}(\beta=1) \simeq 22.4 - 16(n+1.5),
\ee
\be
C^A_{FF}(\beta=1) \simeq 12.2 + 5(n+1.5).
\ee
For the shot-dominated case, the numbers are smaller:
\be
C^B_{\beta\beta}(\beta=1) \simeq 10.8 - 6(n+1.5),
\ee
\be
C^B_{FF}(\beta=1) \simeq 2.7 + (n+1.5).
\ee
We see that, although $C_{\beta\beta}$ is smaller than $C_{FF}$
when $\beta$ is small, $\beta^2 C_{\beta\beta}$ increases
quite significantly with $\beta$, whereas $\beta^2 C_{FF}$ increases
slowly or even declines.
If $\beta$ is large, it is therefore possible to measure $F$
more precisely than $\beta$.

The correlations between the parameters are described by the correlation matrix
\be
r_{ij}=C_{ij}/\sqrt{C_{ii}C_{jj}},
\ee
where $C_{ij} = \langle \delta a_i\, \delta a_j \rangle$, and no summation
is implied in $C_{ii}$.
At $\beta=0$, $\beta$ and $F$  are perfectly anticorrelated, with
$r_{\beta F}$ increasing as $\beta$ increases. In the shot-free case,
$r_{\beta F}(\beta=1)\simeq -0.75+0.4[1.5+n]$, whereas
$r_{\beta F}(\beta=1)\simeq -0.85+0.2[1.5+n]$ when shot-noise dominates.
This increase of the anti-correlation between $F$ with $\beta$  for
lower values of $\beta$ can be understood as follows: 
if we expand equation (20) in powers of $\mu$, for small $\beta$ and 
$\epsilon \equiv F-1$, the coefficient of $\mu^2$ is linear in $\beta$ and 
$\epsilon$, but gives degenerate information on these parameters. 
To separate $\beta$ and $\epsilon$ requires at least the $\mu^4$ term, 
which is of second-order smallness in $\beta$ and $\epsilon$.

These results all apply to long wavelengths, where the effects of the
damping term are negligible. We give in Appendix C a selection
of correlation matrices for this case, and also for the case
where modes up to $k\sigma_p=1$ are used, and $\sigma_p$ is
allowed to float as a further parameter. This makes rather
little difference to the precision with which $F$ can
be determined (assuming the correctness of the model, of course).
The long-wavelength results are illustrated in Figs 5 \& 6,
which show likelihood contours in the $(\beta, F)$ plane where,
at each $(\beta,F)$ point,  the amplitude distribution has been 
integrated over, but the value of $\sigma_p$ is fixed.
These are thus not just slices through the likelihood contours,
but show the practical joint confidence region on $F$ \& $\beta$.

\subsection{Evolutionary effects}

In practice, measurable deviations from $F=1$ 
will involve working over a significant range of redshifts where 
one may need to consider evolution of $F$, $\beta$ etc.
A practical means of dealing with this difficulty will be
to divide the survey up into cubes, and use 
the zero-redshift value of $\ov$ as a parameter.
In each redshift shell, one should form the marginal distribution
for $\ov$, integrating over other parameters as necessary,
before combining all shells to estimate $\ov$ -- in a way that
is by construction independent of evolution
(an alternative would be to fit a functional form $\beta[z]$).
It is the independence of the estimate of $\Lambda$ on evolution which 
makes this method so attractive, as most other geometric tests of $\Lambda$ 
require assumptions about evolution.
In order to apply the Kaiser approximation, we
require subsamples which subtend small angles on the sky;
cubic subsamples will therefore also have a small $\Delta z/(1+z)$
and so it will be satisfactory to ignore evolution within each cube.
The number of modes out to wavenumber $k_{\rm max}$ scales as
the box volume, so the number of modes studied is the same
whether one large volume is studied, or many small sub-volumes.
All that is lost is the signal from the very large-scale
modes, but these are a negligible fraction of the total.

\section{APPLICATIONS}

\subsection{Galaxies}

The next few years should see a number of substantial new galaxy
redshift surveys with which this test could be attempted.
The IRAS PSCz survey of 15,000 galaxies (Saunders et al. 1995)
will cover essentially $4\pi$ sr and will define the density
field out to approximately $z\simeq 0.1$. This will 
not be likely to be so useful for our present purpose owing
both to the high level of shot noise over such
volumes and also to the low depth, so that $F\ls 1.05$ is expected.
More promising is the Sloan survey of $\sim 10^6$ galaxies over
$\pi$ sr, which should provide a map free of shot noise
out to $z\simeq 0.2$, or an effective volume of about 500$^3$
$(h^{-1}\rm Mpc)^3$ (Gunn \& Weinberg 1995). The AAT 2df
facility (Taylor 1995) should provide a survey of slightly greater
depth and approximately $1/6$ the area, giving errors approximately twice
as large as for the Sloan survey.

We have calculated likelihood contours for $\ov$ for a model survey
similar to the 2df survey assuming the same parameter values as in 
section 4.2, which is shown in Fig. 7. The survey is divided into 
cubes each subtending $13^{\circ}$ -- side 100 $h^{-1}$ Mpc at $z = 0.2$ -- the 
number of galaxies in each cube being derived from the APM selection function of 
Baugh \&\ Efstathiou (1993) down to a magnitude limit of $B_J \leq 19.5$ (a surface 
density of approximately 150 per square degree).
We have assumed for simplicity that $\beta$ and $\sigma_{\rm p}\pr$ do not vary 
with redshift; as discussed in section 4.2, this does not affect the
precision in $\ov$ that can be obtained.

\begin{figure}
\japfigC{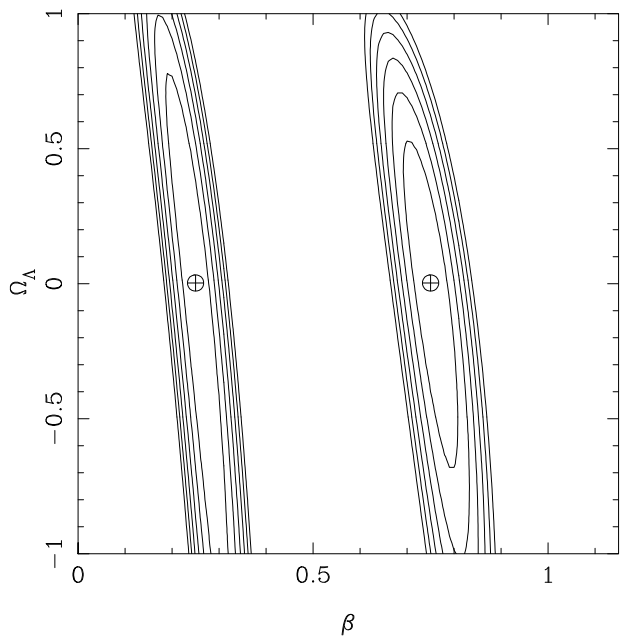}
\caption{Expected contours of likelihood in the $\ov-\beta$ plane
for a 2df-type galaxy redshift survey of 250,000 galaxies to $B_J<19.5$, 
obtained by combining cubic subsamples over a range of redshifts.
Spatial flatness ($\om+\ov=1$) is assumed.
The models shown assume
true values for $\beta$ of 0.25 and 0.75, and $\Omega_\Lambda=0$,
as indicated by the crosses.
}
\end{figure}

We see that the 2df survey is potentially capable of ruling out the high
values of $\ov\simeq0.8$ that are interesting for inflationary models.
Much will depend on the true value of $\beta$ (which will also
be obtained, with exquisite precision). If $\beta\simeq 1$, the rms
error on $\ov$ is about 0.35, but this approximately doubles if $\beta\simeq 0.3$.
In this latter case, even the full Sloan survey would barely suffice
to detect the flattening effect.
This is an ironic outcome, since confusion with the $\beta$ effect is
one of the principal difficulties in measuring the geometrical
flattening. Nevertheless, the best hope for a positive geometrical
detection comes precisely when the redshift-space distortions are highest.

\subsection{Quasars}

Again, both the Sloan and 2df projects are expected to provide large
quasar samples: about $10^5$ and 30,000 respectively.
The larger redshifts probed mean that the anisotropy
signal is expected to be significantly larger, and the
volumes probed are larger, giving many more modes.
However, the quasar samples will be strongly shot-noise limited,
reflecting the low space densities of these tracers.
In Fig. 8 we show results for 30,000 quasars over an area of 750 deg$^2$, 
with a 
constant $dN/dz$ over the redshift range $0.3 \leq z \leq 2.2$. 
For a given value of $\beta$, quasars put tighter constraints on a 
positive $\ov$ than galaxies.  The larger volume, and in
particular the higher redshift, more than offset the increased shot noise, 
especially given that quasars are observed to be 
highly clustered at high redshift : -- the clustering scale-length is 
claimed to be $r_{0} \simeq 6\hMpc$ for quasars
with $z < 2.2$ (Shanks \& Boyle 1994). 

For a particular $\beta$, the proposed quasar surveys thus have 
a strong advantage over the galaxy 
survey. However, $\beta$ for quasars may well be smaller.
The observed quasar clustering corresponds to $\sigma_{8}
\simeq 1.2$ at a mean redshift of $z \simeq 1.5$, whereas it is believed 
that the
present value of $\sigma_{8}$ for mass is about $0.6\om^{-0.5}$; at 
redshift 1.5 this
would be smaller by 2.5 times the $\Omega$-dependent growth suppression factor
(the exact form of which depends on both $\om$ and $\ov$). This 
increases the $\om$ dependence giving roughly:
\bd
\sigma_8(z=1.5) \simeq 0.24\; \om^{-0.8},
\ed
approximately independent of $\Lambda$.
If the ratio of the two $\sigma_8$ values is taken to measure the bias factor,
then quasars are strongly biased at high z and the corresponding $\beta$ is
very low unless $\om$ is extremely low:
\bd
\beta_Q(z=1.5) \simeq 0.2\; \om^{-0.2}
\ed
However, inspection of Fig. 8a shows that, even if $\beta$ for the quasars is
low, they appear to be a  better prospect for ruling out high-$\ov$ 
models than the low-redshift galaxies.     These calculations show that the
advantage of having a larger signal at high redshift outweighs the twin 
disadvantages of higher shot noise and the (anticipated) lower value of 
$\beta$.
However, since the datasets are independent,
the strongest result would come from combining the likelihoods for the two 
surveys.

We have assumed that $\sigma_{\rm p}\pr = 350\; {\rm kms}^{-1}$, 
as for the galaxy survey.
If $\sigma_{\rm p}\pr$ were as low as $200\; {\rm kms}^{-1}$ 
(and this number should certainly decline with redshift, since it
scales roughly linearly with $\sigma_8$)
then $k_{\rm max} =0.8\kMpc$ can be used 
($k_{\rm max}\sigma_{\rm p}\pr = 1.6$). 
However, the constraints on $\ov$ are little different from those of 
Fig. 8 -- the power spectrum amplitude falls off towards small 
scales, hence the additional modes are swamped by the shot noise.

\begin{figure}
\japfigC{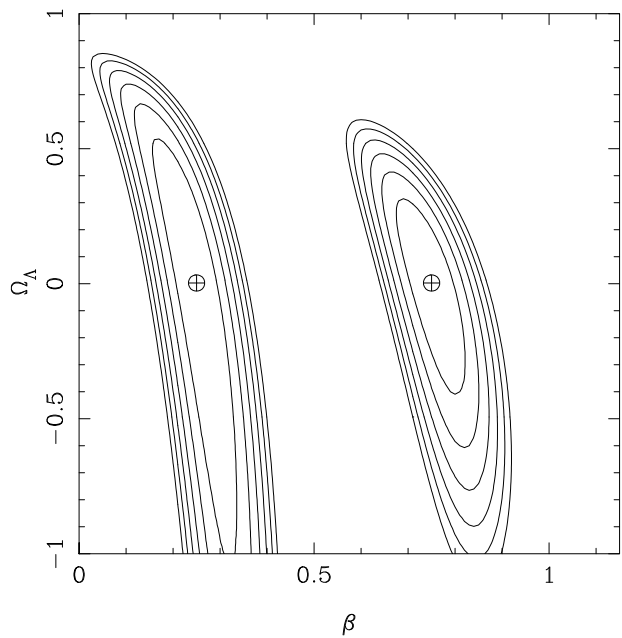}
\caption{Expected contours of likelihood in the $\ov-\beta$ plane 
for a 2df-type quasar survey of 30,000 quasars with $z<2.2$. 
Spatial flatness ($\om+\ov=1$) is assumed.
The models shown assume
true values for $\beta$ of 0.25 and 0.75, and $\Omega_\Lambda=0$,
as indicated by the crosses.
}
\end{figure}

\section{SUMMARY AND CONCLUSIONS}

We have investigated in some detail the practical applicability
of the original suggestion by Alcock \& Paczy\'{n}ski (1979) that
the cosmological constant might be measured via geometrical
distortion of clustering at high redshift.  We find that
application of this method will be considerably more difficult
than previous studies have suggested, for two reasons.

(1) The expectation that $\om\gs 0.2$ limits the likely
degree of anisotropy to a factor $<1.3$, at least for the
spatially flat models popular from inflation.

(2) Without datasets of very large size and quality,
it is difficult to distinguish the geometrical distortion
from the redshift-space anisotropies induced by peculiar
velocities.

In order to have a chance of distinguishing the two
distortions, we have shown that it will be necessary to probe
down to scales where the linear analysis of velocity-induced
anisotropies is inadequate. We have used an approximate model
for such effects to show that datasets likely to become available
in the next 5 years stand a good chance of being able to detect
$\Lambda$.   The exact probability of success depends critically
on the degree of redshift-space distortion, with stronger
signals expected if $\beta$ is high. For $\beta=1$, the
AAT-2df galaxy redshift survey should be able to detect
$\ov\simeq 0.8$, whereas the full Sloan survey would only just
suffice if $\beta\simeq 0.2$. Quasar surveys stand a better
chance of success, provided $\beta_Q$ is not $\ll\beta_G$.
We may expect that surveys being completed now will shortly
settle the controversy over at least $\beta_G$, giving a clearer
idea of how well a geometrical search for $\Lambda$ will work.

Although we have taken a pessimistic tone, showing that
a test for $\Lambda$ which is purely geometrical will
be challenging, there are more optimistic aspects
to our conclusions. The forthcoming surveys will give
extremely precise measurements of $\beta(z)$, and this
will certainly be a useful constraint: a given assumed
$\Lambda$ will then yield $b(z)$, and consistency with the
evolution of the clustering amplitude with redshift should
allow us to rule out some of these possible bias histories.
However, the whole idea of Alcock \& Paczy\'{n}ski's
original suggestion was to avoid dealing with messy
astrophyics, and we have shown that such a `pure'
test is still possible.
This test will require a careful understanding of systematics in
the redshift surveys, but we believe that it can be made into
a reliable method for detecting large vacuum densities.
At present, there is no competing method for estimating $\Lambda$
in a way free from evolutionary uncertainties, so this
route will continue to merit detailed scrutiny.

\subsection*{Acknowledgements}

WEB acknowledges the support of a PPARC research studentship.
Computations were done using STARLINK facilities.

\bigskip
\noindent{\bf REFERENCES}

\bib \strut

\bib Alcock C., Paczy\'{n}ski B., 1979 {\na}, 281, 358

\bib Baugh C.M., Efstathiou G., 1993, {\mn}, 265, 145

\bib Bucher M., Goldhaber A.S., Turok N., 1995, {Phys. Rev. D}, 52, 3314

\bib Carroll S.M., Press W.H., Turner L.E., 1992, ARA{\&}A, 30, 499

\bib Cole S., Fisher K.B., Weinberg D.H., 1994, {\mn}, 267, 785

\bib Cole S., Fisher K.B., Weinberg D.H., 1995, {\mn}, 275, 515

\bib Davis M., Peebles P.J.E., 1983, {\apj}, 267, 465

\bib Efstathiou G., Sutherland W., Maddox S.J., 1990, {\na}, 384, 705

\bib Feldman H.A., Kaiser N., Peacock J.A., 1994, {\apj}, 426, 23

\bib Gunn J., Weinberg, D., 1995,
in {\it Wide-field spectroscopy and the distant universe}, proc. 35th Herstmonceux conference,
eds S.J. Maddox \& A. Arag\'on-Salamanca, World Scientific,
p3.

\bib Heavens A.F., Taylor A.N., 1995, {\mn}, 483, 497

\bib Kaiser N., 1987, {\mn}, 227, 1

\bib Lahav O., Lilje P.B., Primack J.R., Rees M.J., 1991, \mn, 251, 128

\bib Matsubara T., Suto Y., 1996, \apj, in press

\bib Nusser A., Davis M., 1994, {\mn}, 421, L1

\bib Peacock J.A., Nicholson D., 1991, \mn, 253, 307

\bib Peacock J.A., Dodds S.J., 1994, {\mn}, 267, 1020

\bib Peacock J.A., 1992, in Martinez V., Portilla M., S\'aez D., eds, New insights into the Universe, Proc. Valencia summer school (Springer, Berlin), p1

\bib Peebles P.J.E., 1980, The Large Scale Structure of the Universe.
Princeton University Press, Princeton

\bib Phillipps S., 1994, {\mn} 269, 1077

\bib Ryden B., 1995, ApJ, 452, 25

\bib Saunders W., et al., 1995, 
in {\it Wide-field spectroscopy and the distant universe}, proc. 35th Herstmonceux conference,
eds S.J. Maddox \& A. Arag\'on-Salamanca, World Scientific,
p88.

\bib Shanks T., Boyle B.J., 1994, {\mn}, 271, 753

\bib Taylor, K., 1995,
in {\it Wide-field spectroscopy and the distant universe}, proc. 35th Herstmonceux conference,
eds S.J. Maddox \& A. Arag\'on-Salamanca, World Scientific,
p15.

\bib Weinberg S., 1972,  {\it Gravitation \& Cosmology},  (New York: Wiley)

\bib Weinberg S., 1989,  {Rev. Mod. Phys.}, 61, 1

\bib Zaroubi S., Hoffman Y., 1996, Apj, 462, 25

\appendix 

\section{Fourier Transforms}

The squashing effect transforms coordinates and wave vectors as follows:


\be
{\bf x}\pr = S \cdot {\bf x} \;\;\; ; \;\;\; {\bf k}\pr = S^{-1} \cdot {\bf k},
\ee
with transformation matrix

\bd
S = \left( \begin{array}{ccc}
               \fpp^{-1} & 0 & 0 \\
               0 & \fpp^{-1} & 0 \\
               0 & 0 & \fpl^{-1}
             \end{array}
      \right).
\ed
The $\Lambda$ squashing effect on the power spectrum is thus
the following transform of the (possibly anisotropic) correlation function;
\bd
P(\bk) = {\int \xi({\bf r})\; e^{i\smk \cdot \smr}d^{3}r}
\ed
\bea
P^{\prime}(\bk^{\prime}) = {\int \xi({\bf r})\; e^{i(S\cdot\smk\pr) \cdot \smr}\; |S|\;
d^{3}r}
= |S|\; P(S \cdot \bk\pr).
\eea
Note that $\xi\pr({\bf r}\pr) = \xi({\bf r})$ since $\delta\pr(
{\bf r}\pr) = \delta({\bf r})$.
Apart from an amplitude shift the original power spectrum is retained, but
evaluated at $S \cdot \bk\pr$:

\be
P\pr(k_{\parallel}\pr,{\bf k}_{\perp}\pr)=\frac{1}{f_{\perp}^{2}f_{\parallel}}
P\left(\frac{k_{\parallel}\pr}{f_{\parallel}},
  \frac{{\bf k}_{\perp}\pr}{f_{\perp}}\right).
\ee

Redshift distortions are modelled as a product of a Kaiser factor
and a damping term

\bea
P(k_{\parallel},{\bf k}_{\perp})
& = & P_{0}(k) \; (1+\beta\mu^{2})^{2} \; D(k\mu\sigma_{\rm p}) \nn
& = & P_{0}(k) \; k^{-4} \; [k_{\perp}^{2}+(\beta+1)k_{\parallel}^{2}] \;
D(k_{\parallel}\sigma_{\rm p}),
\eea
where $D(k\mu\sigma_{\rm p})$ is the nonlinear `finger of God' correction.
Combining redshift and $\Lambda$ effects gives

\bea
P\pr({\bf k}\pr) & = & \frac{1}{f_{\perp}^{2}f_{\parallel}} \; P_{0}\left({\sqrt
{\frac{\kpp^{\prime 2}}{\fpp^{2}}+\frac{\kpl^{\prime 2}}{\fpl^{2}}}} \; \right)
\left(\frac{\kpp^{\prime 2}}{\fpp^{2}}
+\frac{\kpl^{\prime 2}}{\fpl^{2}}\right)^{-2}  \nn
& \times & \left[\frac{\kpp^{\prime 2}}{\fpp^{2}}+(\beta+1)\frac{\kpl^{\prime
2}}{\fpl^{2}}\right] \;
D\left(\frac{k_{\parallel}\sigma_{\rm p}}{\fpl}\right).
\eea
Now we set $F = \fpl/\fpp$, $\kpl\pr = \mu\pr k\pr$ and $\sigma_{\rm p}\pr =
\sigma_{\rm p}/\fpl$, which yields

\bean
P\pr({\bf k}\pr) & = & \frac{1}{f_{\perp}^{2}f_{\parallel}} \;
P_{0}\left[\frac{k\pr}{\fpp}
\sqrt{1+\mu^{\prime 2}\left(\frac{1}{F^2}-1\right)}\right]\nonumber \\
& \times &
\left[1+\mu^{\prime 2}\left(\frac{1}{F^2}-1\right)\right]^{-2} \nn
& \times &
\left[1+\mu^{\prime 2}\left(\frac{\beta+1}{F^2}-1\right)\right]^{2} 
\; D\left(k\pr\mu\pr\sigma_{\rm p}\pr\right).
\eean
For a power spectrum which is locally close to a power law, with index $n = d\ln P / d\ln k$, we have

\bean
P\pr({\bf k}\pr) & = &
\frac{1}{\fpp^{3+n}F} \; P(k\pr) \;
\left[1+\mu^{\prime 2}\left(\frac{1}{F^{2}}-1\right)\right]^{\frac{n-4}{2}}
\nonumber \\
 & \times& \left[1+\mu^{\prime
2}\left(\frac{\beta+1}{F^{2}}-1\right)\right]^{2}
D\left(k\pr\mu\pr\sigma_{\rm p}\pr\right) 
\eean
Assuming an exponential distribution for the random pairwise velocity
component, the nonlinear correction to the power spectrum is a Lorentzian
factor:

\bd
D\left(k\pr\mu\pr\sigma_{\rm p}\pr\right) =
\frac{1}{1+\left(k\pr\mu\pr\sigma_{\rm p}\pr
\right)^{2}/2}
\ed
(where the units of $\sigma_{\rm p}\pr$ are $h^{-1}\rm Mpc$).

\section{Selection functions and Weighting schemes}

The number of independent modes depends on the details of the survey.
We derive here the effective density of states for a varying selection function
and weighting of the data.
The simplest case is a uniformly sampled cube of side $L$,
containing $N_g$ galaxies, where
independent modes are separated by $\Delta k=2\pi/L$ in $k$ space, so that:
\be
\rho(k) = \left(\frac{L}{2\pi}\right)^{3}.
\ee
In this case (e.g. Peacock \& Nicholson 1991),
one would define Fourier coefficients via
\be
\delta_\bk={1\over N_g}\sum \exp[i{\bf k\cdot x}],
\ee
and in the continuum limit we would have
\be
\langle L^3\; |\delta_\bk|^2 \rangle = P_{\rm true} + P_{\rm shot}
\ee
where the shot noise depends on the density $n=N_g/L^3$:
\be
P_{\rm shot}={1\over n}.
\ee
In these expressions, we use the Fourier convention of Peebles (1980),
but with unit normalization volume:
\be
P(k)=\int \xi(r)\; \exp[i{\bf k\cdot r}]\; d^3r.
\ee
These relations show that there is some advantage to having a large survey,
thus giving a higher density of states. However, for a fixed number of galaxies,
increasing the volume increases the shot noise on each mode. There is an
optimum survey size for a given number of galaxies which corresponds to a
signal-to-noise ratio of about unity.

The general case of a survey of non-uniform sampling was considered
by Feldman et al. (1994; hereafter FKP).
They also allow the galaxy density field to be weighted
so as to optimise the signal-to-noise in the power spectrum:
\be
\delta_\bk={\int w[n - \bar n]\; \exp[i{\bf k\cdot r}]\; d^3r \over
 [\int w^2\bar n^2\; d^3 r]^{1/2} },
\ee
where $\bar n({\bf r})$ is the mean density defined by the sample selection.
With this generalization,
\be
\langle  |\delta_\bk|^2 \rangle = P_{\rm true} + P_{\rm shot}
\ee
where the shot noise is now
\be
P_{\rm shot}={ \int w^2 \bar n\; d^3 r \over
\int w^2 \bar n^2\; d^3 r} 
\ee
FKP also derive (their equation 2.3.2) the ratio of the variance in the
total power to $P^2_{\rm true}$. 
For a uniform cubical survey of side $L$, this would be
\be
{\sigma(P)\over P_{\rm tot} } = N_{\rm modes}^{-1/2}= [\rho_k V_k]^{-1/2},
\ee
which is proportional to $L^{-3/2}$.
Converting the FKP result to the ratio of
total variance to total power gives an effective survey size of
\be
L_{\rm eff}^{-3}={ \int w^4\bar n^4[P+\bar n^{-1}]^2\; d^3 r \over
[P\int w^2\bar n^2\; d^3r + \int w^2\bar n\; d^3r]^2.}
\ee
This reduces to $L_{\rm eff}=L$ for the previous case of a uniform cube.
In general, we should use the optimal FKP weight $w=[1+\bar n P]^{-1}$,
which gives
\be
L_{\rm eff}^{-3}={ \int w^2\bar n^2\; d^3 r \over
[\int w\,\bar n\; d^3r]^2.}
\ee

\section{Covariance matrices}

We present here some example covariance matrices for the parameters $p_{i}$ 
in the form of parameter standard deviations:

\be
\sigma(p_{i}) = \sqrt{C_{ii}}
\ee

\noindent (no summation implied) and correlation matrices:

\be
r_{ij} = C_{ij}/\sigma(p_{i})\sigma(p_{j})
\ee

\noindent for different combinations of values of $\beta$, slope $n$, and 
shot power to true power ratio at $k_{\rm max}$ $R = P_{S}/P_{T}$. The 
standard deviations are scaled such that $\sigma(\beta) = 1$ and the
number of modes $N$ required to achieve this accuracy is given.  For 
a fixed $R$ the standard deviations scale as $N^{-1/2}$.  
For example, if the volume of the survey 
is increased by a factor of 100, and $k_{\rm max}$ and the number density of 
galaxies were kept constant, the errors on the parameters would decrease 
by a factor of $10$. We show covariances 
for both the three-parameter model with $\sigma_{\rm p}\pr$ fixed and the 
full four-parameter model where it is allowed to vary.   The number of modes
is kept constant when the number of parameters is changed, so the value of
$\sigma(\beta)$ reflects the additional uncertainty introduced.  
All models have $k_{\rm max} \sigma_{\rm p}\pr = 1$ and $F=1$.

\subsection{Example 1: $\beta=1$; $n=-1.5$; $R=0$; $N=1120$}

\bigskip
\vbox{
\noindent{\bf Table C1}. Three-parameter model.\\
\\
\begin{center}
\begin{tabular}{r|r|r|r}
  & $\beta$ & $F$ & $\ln A$\\
\\
 $\sigma(p)\;\;\;$ & 1.00 & 0.74 & 0.43 \\
\\
 $\beta\;\;\;$ & 1.00 & $-$0.76 & $-$0.86 \\
 $F\;\;\;$ & $-$0.76 & 1.00 &  0.56 \\
 $\ln A\;\;\;$ & $-$0.86 & 0.56 & 1.00
\\
\end{tabular}
\end{center}
}

\bigskip
\vbox{
\noindent{\bf Table C2}. Four-parameter model.\\
\\
\begin{center}
\begin{tabular}{r|r|r|r|r}
  & $\beta$ & $F$ & $\sigma_{\rm p}\pr$ & $\ln A$\\
\\
 $\sigma(p)\;\;\;$ & 1.38 & 0.75 & 2.52 & 0.42 \\
\\
 $\beta\;\;\;$ & 1.00 & $-$0.43 & 0.69 & $-$0.63 \\
 $F\;\;\;$ & $-$0.43 & 1.00 & 0.16 & 0.55 \\
 $\sigma_{\rm p}\pr\;\;\;$ & 0.69 & 0.16 & 1.00 & $-$0.01 \\
 $\ln A\;\;\;$ & $-$0.63 & 0.55 & $-$0.01 & 1.00
\\
\end{tabular}
\end{center}
}

\subsection{Example 2: $\beta=0.3$; $n=-1.5$; $R=0$; $N=3210$}

\bigskip
\vbox{
\noindent{\bf Table C3}. Three-parameter model.\\
\\
\begin{center}
\begin{tabular}{r|r|r|r}
  & $\beta$ & $F$ & $\ln A$\\
\\
 $\sigma(p)\;\;\;$ & 1.00 & 1.14 & 0.24 \\
\\
 $\beta\;\;\;$ & 1.00 & $-$0.96 & $-$0.73 \\
 $F\;\;\;$ & $-$0.73 & 1.00 &  0.58 \\
 $\ln A\;\;\;$ & $-$0.73 & 0.58 & 1.00
\\
\end{tabular}
\end{center}
}

\bigskip
\vbox{
\noindent{\bf Table C4}. Four-parameter model.\\
\\
\begin{center}
\begin{tabular}{r|r|r|r|r}
  & $\beta$ & $F$ & $\sigma_{\rm p}\pr$ & $\ln A$\\
\\
 $\sigma(p)\;\;\;$ & 1.09 & 1.14 & 1.49 & 0.24 \\
\\
 $\beta\;\;\;$ & 1.00 & $-$0.87 & 0.41 & $-$0.66 \\
 $F\;\;\;$ & $-$0.87 & 1.00 & $-$0.01 & 0.59 \\
 $\sigma_{\rm p}\pr\;\;\;$ & 0.41 & $-$0.01 & 1.00 & 0.00 \\
 $\ln A\;\;\;$ & $-$0.66 & 0.59 & 0.00 & 1.00
\\
\end{tabular}
\end{center}
}

\subsection{Example 3: $\beta=1$; $n=-1$; $R=0$; $N=715$}

\bigskip
\vbox{
\noindent{\bf Table C5}. Three-parameter model.\\
\\
\begin{center}
\begin{tabular}{r|r|r|r}
  & $\beta$ & $F$ & $\ln A$\\
\\
 $\sigma(p)\;\;\;$ & 1.00 & 1.01 & 0.53 \\
\\
 $\beta\;\;\;$ & 1.00 & $-$0.57 & $-$0.86 \\
 $F\;\;\;$ & $-$0.57 & 1.00 &  0.56 \\
 $\ln A\;\;\;$ & $-$0.86 & 0.56 & 1.00
\\
\end{tabular}
\end{center}
}

\bigskip
\vbox{
\noindent{\bf Table C6}. Four-parameter model.\\
\\
\begin{center}
\begin{tabular}{r|r|r|r|r}
  & $\beta$ & $F$ & $\sigma_{\rm p}\pr$ & $\ln A$\\
\\
 $\sigma(p)\;\;\;$ & 1.59 & 1.03 & 3.13 & 0.53 \\
\\
 $\beta\;\;\;$ & 1.00 & $-$0.23 & 0.78 & $-$0.55 \\
 $F\;\;\;$ & $-$0.23 & 1.00 & 0.16 & 0.55 \\
 $\sigma_{\rm p}\pr\;\;\;$ & 0.78 & 0.16 & 1.00 & $-$0.01 \\
 $\ln A\;\;\;$ & $-$0.55 & 0.55 & $-$0.01 & 1.00
\\
\end{tabular}
\end{center}
}

\subsection{Example 4: $\beta=1$; $n=-1.5$; $R=1000$; $N=855000$}

Note that in this shot-noise dominated regime, the standard deviations 
scale with $R$ to a good approximation for fixed $N$.

\bigskip
\vbox{
\noindent{\bf Table C7}. Three-parameter model.\\
\\
\begin{center}
\begin{tabular}{r|r|r|r}
  & $\beta$ & $F$ & $\ln A$\\
\\
 $\sigma(p)\;\;\;$ & 1.00 & 0.51 & 0.52 \\
\\
 $\beta\;\;\;$ & 1.00 & $-$0.85 & $-$0.92 \\
 $F\;\;\;$ & $-$0.85 & 1.00 &  0.65 \\
 $\ln A\;\;\;$ & $-$0.92 & 0.65 & 1.00
\\
\end{tabular}
\end{center}
}

\bigskip
\vbox{
\noindent{\bf Table C8}. Four-parameter model.\\
\\
\begin{center}
\begin{tabular}{r|r|r|r|r}
  & $\beta$ & $F$ & $\sigma_{\rm p}\pr$ & $\ln A$\\
\\
 $\sigma(p)\;\;\;$ & 1.02 & 0.51 & 1.35 & 0.52 \\
\\
 $\beta\;\;\;$ & 1.00 & $-$0.82 & 0.17 & $-$0.91 \\
 $F\;\;\;$ & $-$0.82 & 1.00 & 0.03 & 0.65 \\
 $\sigma_{\rm p}\pr\;\;\;$ & 0.17 & 0.03 & 1.00 & 0.00 \\
 $\ln A\;\;\;$ & $-$0.91 & 0.65 & 0.00 & 1.00
\\
\end{tabular}
\end{center}
}

\end{document}